\documentclass{mem}
\input psfig.sty
\usepackage{natbib}\usepackage{txfonts}\usepackage{balance}
\usepackage{graphicx}
\usepackage[a4paper]{hyperref}
\idline{80}{ 000 }
\begin{document}

\title{
Constraining Dark Matter-Dark Energy Interaction
with Gas Mass Fraction in Galaxy Clusters
}

   \subtitle{}

\author{
R. \,S. \,Gon\c calves\inst{1}, J. \,S. \,Alcaniz\inst{1}, A. \,Dev\inst{2} and D. \,Jain\inst{3}
}

  \offprints{R. \,S. \,Gon\c calves}

\institute{
Observat\'orio Nacional, Rio de Janeiro - RJ, 20921-400, Brasil\\ 
\email{rsousa@on.br}; \email{alcaniz@on.br}
\and
Miranda House, University of Delhi, Delhi - 110007, India
\and
Deen Dayal Upadhyaya College, University of Delhi, Delhi - 110015, India\\ \email{djain@ddu.du.ac.in}
}

\authorrunning{Gon\c calves}

\titlerunning{$f_{gas}$ constraints on DM-DE interaction}

\abstract{
The recent observational evidence for the current cosmic acceleration have
stimulated renewed interest in alternative cosmologies, such as scenarios with
interaction in the dark sector (dark matter and dark energy). In general, such
models contain an unknown negative-pressure dark component coupled with
the pressureless dark matter and/or with the baryons that results in an evolution
for the Universe rather different from the one predicted by the standard $\Lambda$CDM
model. In this work we test the observational viability of such scenarios by
using the most recent galaxy cluster gas mass fraction versus redshift data ($42$
X-ray luminous, dynamically relaxed galaxy clusters spanning the redshift
range $0.063 < z < 1.063$) \citep{Allen08} to place bounds on the parameter $\epsilon$ that
characterizes the dark matter/dark energy coupling. The resulting are consistent
with, and typically as constraining as, those derived from other cosmological
data. Although a time-independent cosmological constant ($\Lambda$CDM model) is a good fit to these galaxy cluster data, an interacting dark energy
component cannot yet be ruled out.
}

\maketitle{}

\section{Introduction}

Although fundamental to our understanding of the Universe, several important
questions involving the nature of the dark matter and dark energy components
and their roles in the dynamics of the Universe remain unanswered. Among
these questions, the possibility of interaction in the dark sector, which gave
origin to the so-called models of coupled quintessence, has been largely
explored in the literature. These scenarios are based on the premise that, unless
some special and unknown symmetry in Nature prevents or suppresses a non-minimal
coupling between these components (which has not been found), such
interaction is in principle possible and, although no observational piece of
evidence has so far been unambiguously presented, a weak coupling still below
detection cannot be completely excluded (see \citealt{amendola};\citealt{zimd,barrow,wu,ernandes,ernandes1}).

From the observational viewpoint, these models are capable of explaining the
current cosmic acceleration, as well as other recent observational results. From
the theoretical point of view, however, critiques to these scenarios do exist and
are mainly related to the fact that in order to establish a model and study their
observational and theoretical predictions, one needs first to specify a
phenomenological coupling between the cosmic components.

In this work, instead of adopting the traditional approach, we follow the qualitative
arguments used in {\citep{Wang05, Alcaniz05}} and deduce the interacting law from a simple
argument about the effect of the dark energy decay on the dark matter (CDM)
expansion rate. The resulting expression is a very general law that has many of
the previous phenomenological approaches as a particular case~(see, e.g., {{\citealt{Wang05}} for a discussion).

\section{The model}

By assuming that the radiation and baryonic fluids are separately conserved, the energy conservation law for the two interacting components ($u_{\alpha} T^{\alpha\beta}_{;\beta} = 0$, where $T^{\alpha\beta} = T^{\alpha\beta}_{DM} + T^{\alpha\beta}_{DE}$) reads

\begin{equation}
\label{DarkCompCons}
\dot{\rho}_{DM} + 3 \frac{\dot{a}}{a}\left(\rho_{DM} \right) =  - \dot{\rho}_{DE} - 3 \frac{\dot{a}}{a}\left(\rho_{DE} + p_{DE} \right).
\end{equation}
In order to complete the description of our interacting quintessence scenario we
need to specify the interaction law. In principle, if the quintessence component
is decaying into CDM particles, the CDM component will dilute more slowly
compared to its standard (conserved) evolution. Therefore, if the deviation from
the standard evolution is characterized by a constant $\epsilon$ we may write
\begin{equation}
\label{DMEvol}
\rho_{DM} = \rho_{DM,0} a^{-3+\epsilon}\;,
\end{equation}
where the subscript 0 stands for present-day quantities and we have set $a_0 = 1$.

In what follows we also consider that the dark energy component is described
by the equation of state $p = w\rho$, where the constant $w$ is a negative quantity.
Now, by integrating Eq. (1) it is straightforward to show that the energy density
of the dark energy component is given by~~\citep{jesus08}
\begin{equation}
\label{DEEvol}
\rho_{DE} = \tilde{\rho}_{DE,0} a^{-3(1 + w)} + \frac{\epsilon \rho_{DM,0}}{3 \left| w \right| - \epsilon} a^{-3 + \epsilon}\;.
\end{equation}
Clearly, in the absence of a coupling with the CDM component, i.e., $\epsilon = 0$, the conventional non-interacting quintessence scenario is fully recovered. Neglecting the radiation contribution, the Friedmann equation for this interacting dark matter-dark energy cosmology can be written as

\begin{equation}
\label{EqFried}
{\cal{H}}^2 = \Omega_b a^{-3} + \frac{3\left| w \right| \Omega_{DM}}{3 \left| w \right| - \epsilon}a^{-3 + \epsilon} + \tilde{\Omega}_{DE} a^{-3(1 + w)}  \; , 
\end{equation}
where ${\cal{H}} = H(z)/H_0$ and $\Omega_b$ and $\Omega_{DM}$ are, respectively, the baryons and CDM present-day density
parameters.

The parameter $\Omega_{DE}$ is defined, in terms of the density parameter of the dark
energy component as
\begin{equation}
\label{DensParam}
\tilde{\Omega}_{DE} = \Omega_{DE} - \frac{\epsilon \Omega_{DM}}{3 \left| \omega \right| - \epsilon}\;,
\end{equation}
and, therefore,
\begin{equation}
\label{NormCond}
\tilde{\Omega}_{DE} = 1 - \Omega_b - \frac{3 \left| \omega \right| \Omega_{DM}}{3 \left| \omega \right| - \epsilon}\;.
\end{equation}

The above expression clearly shows that the conventional (non-interacting)
quintessence scenario is considerably modified due to the dark energy decay
into CDM particles. It is also worth noticing the importance of the baryonic
contribution to this sort of scenario. Going back to high redshift we see that the
presence of an explicit term redshifting as $(1+z)^3$ is well justified since the
decaying dark energy component slows down the variation of of CDM density.
Although being sub-dominant at the present stage of cosmic evolution, the
baryonic content will be dominant (in comparision to CDM) at very high
redshifts. Actually, it becomes sub-dominant just before nucleosynthesis ($z \sim
10^{10}$ for $\epsilon \sim 0.1$), so that the CDM component drives the evolution after the
radiation phase.

\section{Constraints from X-Ray gas mass fraction} 

In our analysis we consider the Chandra data analyzed in \citep{Allen08}. The
data sample consists of $42$ clusters distributed over a wide range of redshift ($0.063 < z
< 1.063$). The clusters studied are all regular, relatively relaxed systems for
which independent confirmation of the matter density parameter results is
available from gravitational lensing studies. The default cosmology used is the
$\Lambda CDM$ scenario with $H_{0}=70 \, km \, s^{-1}Mpc^{-1}$. By assuming that the baryonic
mass fraction in galaxy clusters provides a fair sample of the distribution of
baryons at large scale \citep{White93}, as well as $f_{gas} \propto {d_A}^{3/2}$ \citep{Sasaki96} (where $d_A$ is the angular diameter distance), the model function is defined by \citep{Allen08}
\begin{equation}
\label{Fgas}
f_{gas}^{mod} (z)= {\cal{G}} \frac{\Omega_{b}}{\Omega_{b} + \Omega_{DM}a^{\epsilon}}\left[ \frac{d^{\Lambda CDM}_{A}(z)}{d^{mod}_{A}(z)} \right]^{\frac{3}{2}}\;,
\end{equation}
where ${\cal{G}} = \frac{K \, A \, \gamma \, b(z)}{1+s(z)}$, the parameters $K$, $A$, $\gamma$, $b(z)$ and $s(z)$ account for computational simulations
for corrections under the assumption that $f_{gas}$ should be approximately constant
with $z$. On the other hand, the distance ratio $[ {d^{\Lambda CDM}_{A}(z)}/{d^{mod}_{A}(z)}]^{{3}/{2}}$ accounts for deviations in the geometry of the universe from the $\Lambda CDM$ model, where explicitly we have ($c = 1$)

\begin{equation}
\label{DAngLCDM}
d^{\Lambda CDM}_A = \frac{1}{1+z} \int^{z}_{0}{\frac{dz}{H_0\sqrt{0.3a^{-3} + 0.7}}}
\end{equation}
and
\begin{equation}
\label{DAngMod}
d^{mod}_A = \frac{1}{1+z} \int^{z}_{0}{\frac{dz}{H_0{\cal{H}}}} 
\end{equation}
where ${\cal{H}}$ is given by Eq. (4).

In order to determine the cosmological parameters $\epsilon$ and $\Omega_{DM}$ we use a $\chi^2$ minimization. For this purpose we set $A = 1.0$ \citep{Samushia08}, as well $\Omega_b = 0.044$~\citep{Spergel93} and marginalize over the other parameters. The expression for $\chi^2$ is:
\begin{equation}
\label{Chi2Geral}
\chi^2 = \sum_{i = 1}^{42} \frac{\left[ f_{gas}^{obs}(z_i ; \textbf{p}) - f_{gas}^{mod}(z_i ; \textbf{p}) \right]^{2}}{\sigma^2_{f^{obs}_{gas}} }\;,
\end{equation}
where $\sigma_{f_{gas},i}$ are the symmetric root-mean-square errors for the $\Lambda CDM$ data. The $68.3\%$ and $95.4\%$ confidence levels are defined by the conventional two-parameters $\chi^2$ levels $2.30$ and $6.17$, respectively.

\begin{table}[t]  
\begin{center}  
\begin{tabular}{rrll}  
\hline  \hline \\
\multicolumn{1}{c}{$\Omega_{DM}$ }&
\multicolumn{1}{c}{$\epsilon$ }& 
\multicolumn{1}{c}{$\chi^2_{min}/\nu$}& 
\multicolumn{1}{c}{$w$}\\  \\  \hline  \hline
\\
$0.207^{+0.010}_{-0.016}$& $-0.215^{+0.167}_{-0.166}$& 1.24& -0.8\\ 
\\
$0.206^{+0.009}_{-0.011}$& $- 0.040^{+0.231}_{-0.180}$& 1.21& -1.0\\
\\
$0.204^{+0.011}_{-0.024}$& $0.120^{+0.177}_{-0.069}$& 1.25 & -1.2\\
\\
\hline  \hline
\end{tabular} 
\caption{Best-fit values of $\Omega_{DM}$ and $\epsilon$.}  
\end{center} 
\end{table} 

Figs. 1, 2 and 3 show contours of constant likelihood ($68.3\%$ and $95.4\%$) in the $\epsilon-\Omega_{DM}$ plane for fixed values of $w = -0.8$ (coupled quintessence), $w = -1$ (vacuum decay) and $w = -1.2$ (coupling with phantom), respectively. The best-fit values and 68.3\% error estimates for the parameters $\Omega_{DM}$ and $\epsilon$ are shown in Table 1. We note that for values of $w \geq -1.0$, the observational analysis favours a transfer of energy from dark matter to dark energy ($\epsilon < 0$). 

%At 68.3\% c.l. we find $\Omega_{DM} = 0.207^{+0.010}_{-0.016}$ and $\epsilon = -0.215^{+0.167}_{-0.166}$ (w = -0.8); $\Omega_{DM} = 0.206^{+0.009}_{-0.011}$ and $\epsilon = - 0.040^{+0.231}_{-0.180}$, with $\chi^2_{min}/\nu = 1.211$ (when $\omega = -1.0$) and $\Omega_{DM} = 0.204^{+0.011}_{-0.024}$ and $\epsilon = 0.120^{+0.177}_{-0.069}$, with $\chi^2_{min}/\nu = 1.255$ (when $\omega = -1.2$). Note that for values of $w \geq -1.0$, the observational analysis favours a transfer of energy from dark matter to dark energy ($\epsilon < 0$).

\begin{figure*}[t]
\centerline{
\psfig{figure=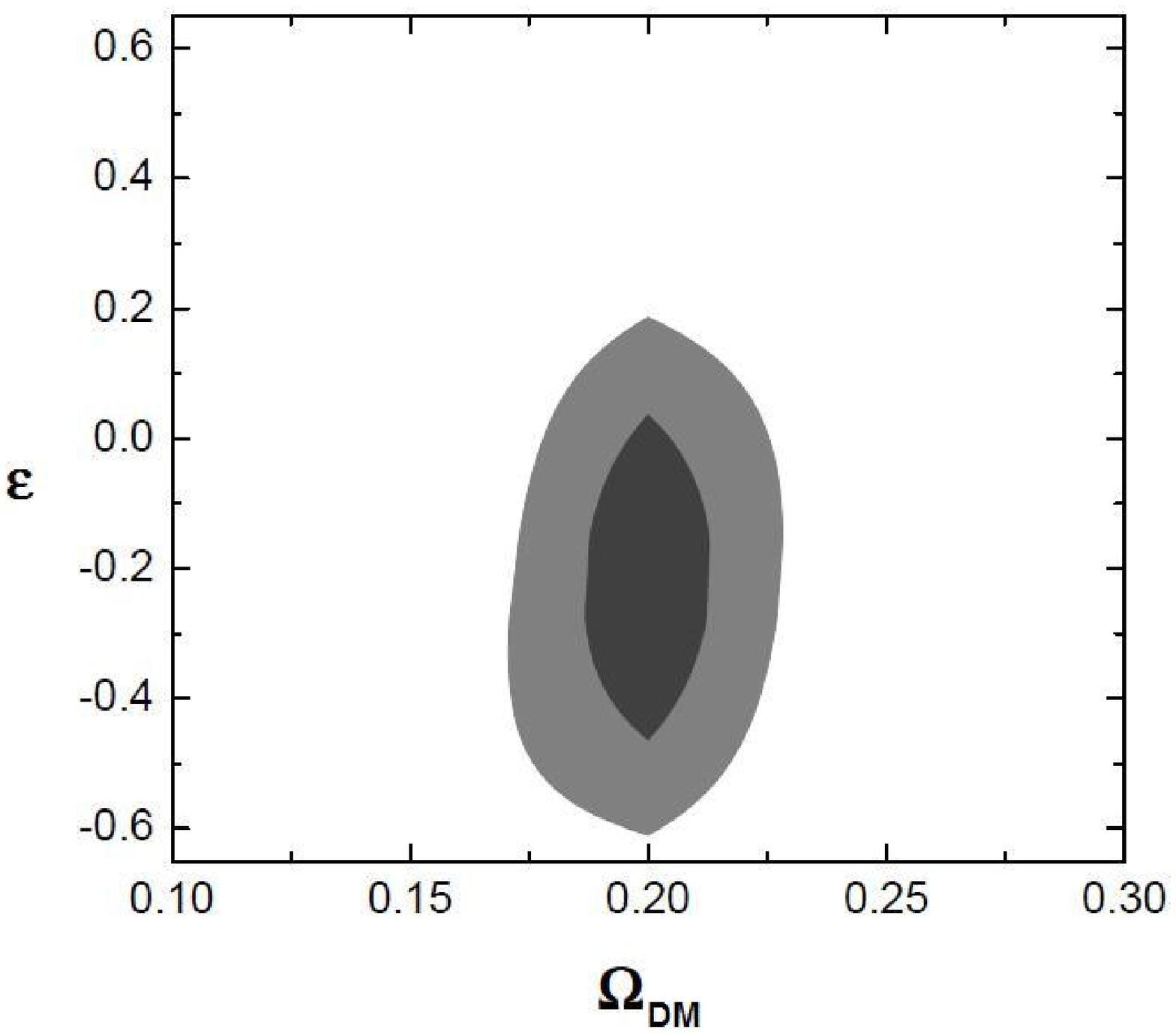,width=2.3truein,height=3.2truein,angle=0}
\hskip -0.95cm
\psfig{figure=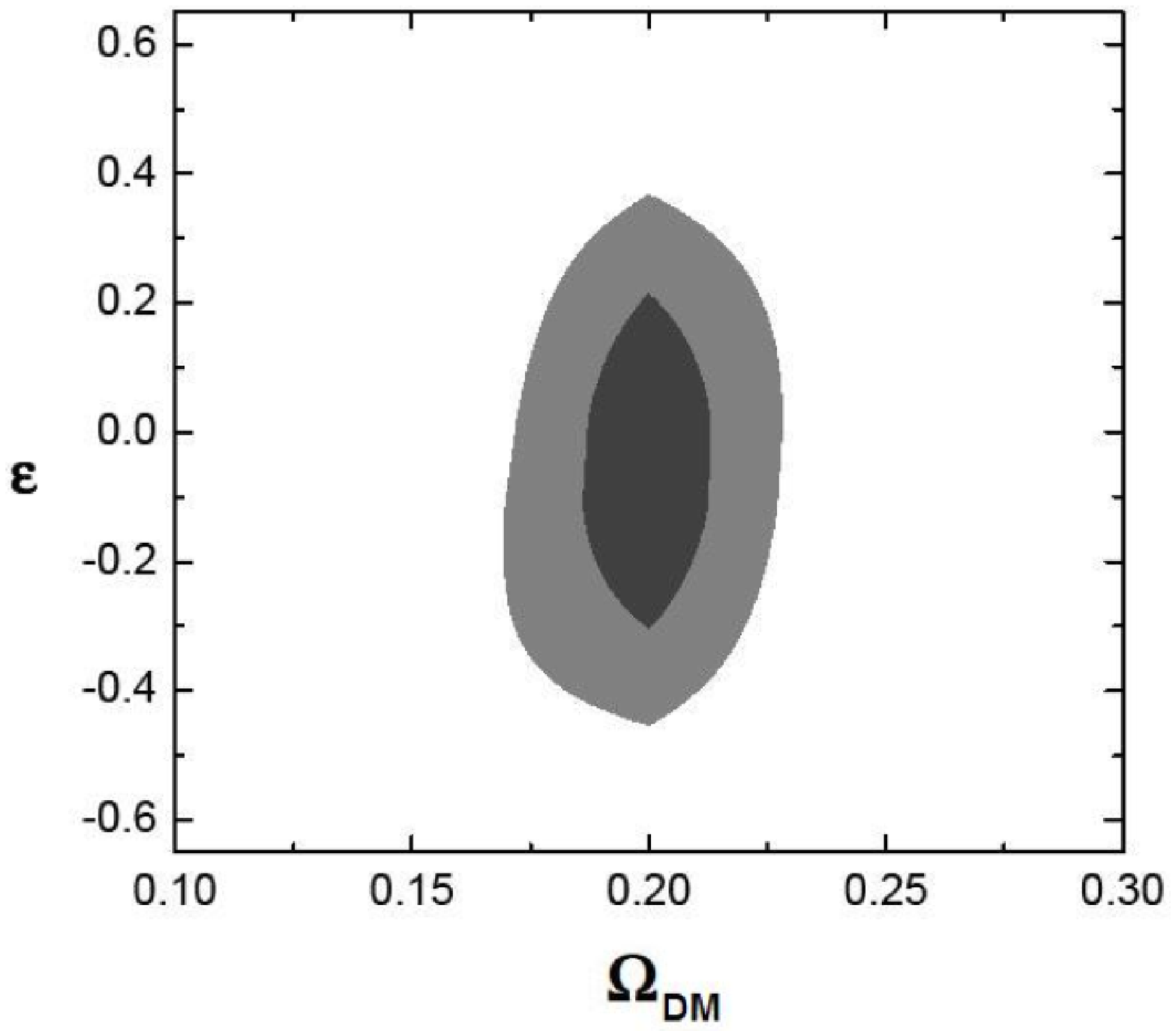,width=2.3truein,height=3.2truein,angle=0}
\hskip -0.95cm
\psfig{figure=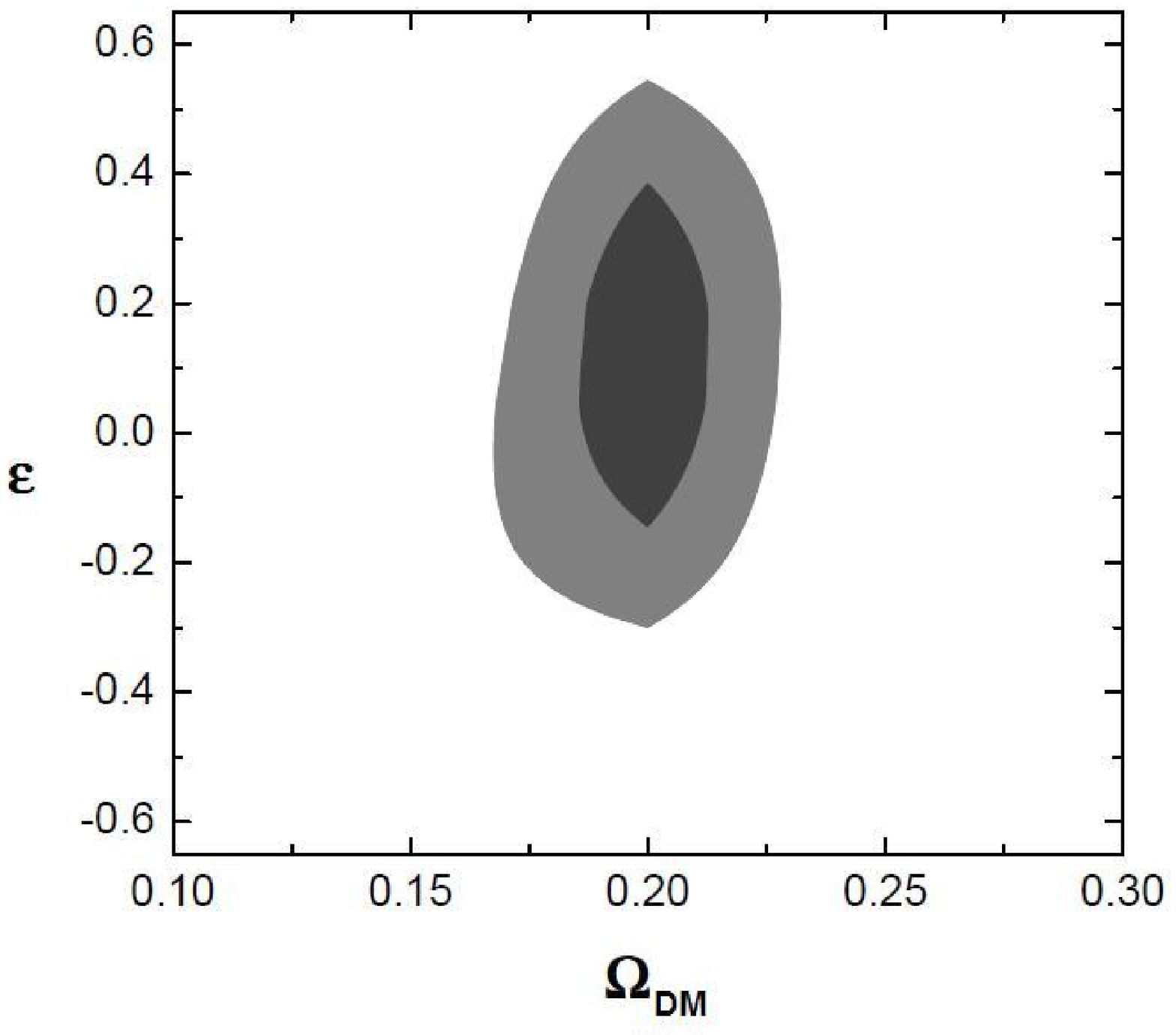,width=2.3truein,height=3.2truein,angle=0}
}
\vskip -1.3cm
\caption{Confidence regions ($68.3\%$ and $95.4\%$) in the $\epsilon - \Omega_{DM}$ plane. From left to right the values of $w$ correspond to -0.8, -1.0 and -1.2, respectively.}
\vskip 1.0cm
\end{figure*}

\section{A low-$z$ test for DM-DE interaction}

From Eq. (8), we see that at low-$z$ the dependence with the cosmological model can be easily eliminated. Thus, we can rewrite the $f_{gas}$ function as
\begin{equation}
\label{FgasZ}
f_{gas}^{mod} (z)= \frac{K \, A \, \gamma \, b(z)}{1+s(z)} \frac{\Omega_{b}}{\Omega_{b} + \Omega_{DM}a^{\epsilon}}\;.
\end{equation}
Note that, although model-independent, the above equation can be used to distinguish between coupled and uncoupled scenarios since the the $a^{\epsilon}$ term cannot be removed. As an example, by restricting our sample to $11$ clusters at $z \leq 0.23$, we perform our $\chi^2$ analysis and find $\epsilon = -0.065$. A more detailed discussion on this low-$z$ test involving  a new set of $f_{gas}$ data will appear in a forthcoming communication~\citep{rodrigo}.

\section{Conclusions}

We have used gas mass fraction data to test the viability of a general class of coupled quintessence models. In particular, we have proposed a new low-$z$ test which in principle may be able to distinguish classes of coupled and uncoupled dark matter/energy models. We believe that with new and more
precise data sets we will be able to use this test to place tighter and stronger bounds on the interacting parameter $\epsilon$.

\bibliographystyle{aa}

\end{document}